\documentclass[aip,
reprint,%
]{revtex4-1}
\usepackage[T1]{fontenc}
\usepackage[utf8]{inputenc}
\usepackage{graphicx}
\usepackage{amsmath}
\usepackage{amssymb}
\usepackage{color} 
\usepackage{pifont}
\usepackage{natbib}
\usepackage[left,columnwise]{lineno}
\usepackage{xcolor}
\newcommand{\jcor}{\textcolor{black}}

\begin{document}

\thanks{\textsuperscript{*} These authors contributed equally.}

\title{Local and Global Expansivity in Water}
\author{Jan Philipp Gabriel\normalfont{\textsuperscript{*}}}
\email[Correspondence to: ]{jan.gabriel@dlr.de}
\affiliation{Institute of Materials Physics in Space, German Aerospace Center, 51170 Köln, Germany}

\author{Robin Horstmann}
\affiliation{Institute for Condensed Matter Physics, Technical University Darmstadt, 64289 Darmstadt, Germany}

\author{Martin Tress\normalfont{\textsuperscript{*}}}
\email[Correspondence to: ]{martin.tress@uni-leipzig.de \\ \textsuperscript{*} These authors contributed equally.}
\affiliation{Peter Debye Institute for Soft Matter Research, Leipzig University, 04103 Leipzig, Germany}

\date{\today} 

\begin{abstract}
\jcor{The supra-molecular structure of a liquid is strongly connected to its dynamics which in turn controls macroscopic properties such as viscosity. Consequently, detailed knowledge about how this structure changes with temperature is essential to understand the thermal evolution of the dynamics ranging from the liquid to the glass. Here we combine infrared spectroscopy (IR) measurements of the hydrogen (H) bond stretching vibration of water with molecular dynamics simulations and employ a quantitative analysis to extract the inter-molecular H-bond length in a wide temperature range of the liquid. The extracted expansivity of this H-bond differs strongly from that of the average nearest neighbor distance of oxygen atoms obtained through a common conversion of mass density. However, both properties can be connected through a simple model based on a random loose packing of spheres with a variable coordination number which demonstrates the relevance of supra-molecular arrangement. Further, the exclusion of the expansivity of the inter-molecular H-bonds reveals that the most compact molecular arrangement is formed in the range $\sim316-331\,\text{K}$ (i.e. above the density maximum) close to the temperature of several pressure-related anomalies which indicates a characteristic point in the supra-molecular arrangement. These results confirm our earlier approach to deduce inter-molecular H-bond lengths via IR in polyalcohols [Gabriel et al. J. Chem. Phys. 154, 024503 (2021)] quantitatively and open a new alley to investigate the role of inter-molecular expansion as a precursor of molecular fluctuations on a bond-specific level.}
\end{abstract}
\maketitle 


\section{Introduction}

The macroscopic properties of a liquid like its shear viscosity or its surface tension are controlled by the dynamics, the interactions, and the arrangement of its molecules. However, the particular relations between the macroscopic and the molecular quantities are often obscured by the fact that the molecular properties are interconnected to each other through complicated and in part even unknown relations. In simple model liquids, the dynamics of the molecules have been connected to the mass density of the system \cite{Xia2000,Stevenson2005,Lubchenko2007}. However, many experimental studies of real materials show that such a relation is not unambiguously applicable; especially hydrogen (H) bonding systems cannot be described in this way \cite{Floudas:2010,Pawlus2011,Paluch2014,Hensel:2004}. This indicates, that other factors beyond the average mass density like the particular supra-molecular structure may also affect liquid dynamics on the molecular and thus in turn also on the macroscopic scale. Consequently, a comprehensive understanding of the dynamics and its thermal evolution in the liquid state and also in the glass requires detailed insight into the supra-molecular structure and its changes with temperature \cite{Lunkenheimer2023}.\\
Among all materials water is one of the most-studied liquids in any aspect ranging from (but not limited to) structural investigations by means of X-ray \cite{Gorbaty1985,Narten2003,Head-Gordon2006,Fu2009,Naberukhin2017,Soper2007,Bergmann2007,Amann-Winkel2016} and neutron scattering \cite{Krauss2019,Soper2007,Amann-Winkel2016}, examination of inter-molecular interactions using vibrational spectroscopy \cite{Ford1968,Marechal1991,Corcelli2005,DeMarco2016,Zhou2022}, and also computer simulations testing different classical and quantum mechanical models \cite{Marti1994,Corcelli2005,Choi2013,Nagata2015,Carlson2020,Liu2021a}. Thereby, the coordinated inter-molecular H-bond is the central interaction, and its importance for the various anomalies of the material itself \cite{Gainaru2014,Gallo2021,Kontogeorgis2022} as well as the sophisticated structure formation of biological molecules in aqueous solution \cite{Capaccioli2020,Wolf2015,Amann-Winkel2016} has led to an exceptional research interest.\\
However, for most other liquids no such vast collection of structural and dynamical data as well as theoretical model collection exists. Therefore, we aim to develop a time- and cost-effective approach to obtain deeper insight into the supra-molecular structure of liquids. An earlier attempt has revealed intricate insights into the bond-specific expansivity of polyalcohols and in turn offered a clear explanation for the severed connection between their density and molecular relaxation dynamics \cite{gabriel2021}.\\
To explore the scope of this approach in depth, water represents a well-known reference system with many data collections in the literature for verification. Furthermore, water has a simple molecular structure which rules out any conformational changes. Because of that, the supra-molecular aggregate is exclusively constituted of a network of inter-molecular H-bonds, and the structural parameters are only its length and the orientation of the molecules with respect to each other (which results in the formation of a particular average number of inter-molecular H-bonds per molecule).\\

Here we present an investigation of H-bond stretching vibrations of H$_2$O and HDO diluted in D$_2$O employing experiments and molecular dynamics simulations which both aim at testing whether the correspondingly extracted bond lengths truly can describe macroscopic lengths and expansions. The obtained spectra exhibit pronounced contributions from combined vibrations which are strongly suppressed in dilute HDO where mostly the vibrations isolated OH bonds prevail. Following the implication that only the latter reflects the actual state of the bond, we also draw more general conclusions on the interpretation of the vibrational spectra of H2O in favor of contributions from isolated and combined vibrations instead of splitting into different states characterized by the number of donor and acceptor molecules in the direct vicinity. Furthermore, feeding an elementary model based on a random loose packing of spheres with the obtained bond lengths and the macroscopic mass density yields an average coordination number of $3.5-3.6$ which is well in line with the literature. This confirms our assumption regarding the reliability of molecular length deduced from vibrational spectra and opens a new alley to investigate intermolecular expansion as a precursor of molecular fluctuations on a bond-specific level.\\ 

\section{Materials and Methods}
\subsection*{H$_2$O and HDO in D$_2$O}
MilliQ water with a specific resistivity of $12$\,M$\Omega $cm was used in the H$_2$O experiments. A solution of 2 \% HDO in D$_2$O was prepared by mixing 1 wt\% MilliQ water in D$_2$O (used as received from Sigma Aldrich).

\subsection*{Infrared spectroscopy measurements}
IR spectra were recorded with a Fourier transform infrared (FTIR) spectrometer (Bio–Rad FTS 6000) combined with an IR microscope (UMA 500) and a liquid nitrogen-cooled mercury-cadmiumtelluride (MCT) detector (Kolmar Technologies, Inc, USA) while a closed liquid cell (supplier) was used to control the sample temperature.


\subsection*{Molecular dynamics simulations}
The simulations of water employed the TIP4P/2005f water model \cite{gonzalez2011flexible}, an extension of the very successful TIP4P/2005 water model with a flexible geometry to address IR absorption. Thereby, the intra-molecular OH bond lengths are described by a Morse potential, and a harmonic potential is used for the intra-molecular HOH angle. The parametrization of these potentials leads to good agreement with experimental results for the bending and stretching motions. These vibrations, together with the point charges on the hydrogen atoms and the additional site of the negative charge with relative positions to all three atoms, lead to oscillations of the molecular dipole moment in the IR range of the spectrum.\\
A bulk water system consisting of 2000 molecules was studied by molecular dynamics (MD) simulations using the GROMACS \cite{Drunen_CPC_1995_gromacs, Lindahl_S_2015_gromacs} simulation package (version 2019.4). The temperature was set using the Velocity-Rescaling thermostat \cite{Parrinello_JCP_2007_canonical}, with a time constant of 1.3\,ps, and the density was equilibrated at each temperature at a pressure of 1\,bar with the Parrinello-Rahman barostat \cite{Rahman_JAP_1981_polymorphic} with a time constant of 2.3\,ps and a compressibility of 4.5e-5\,bar. Lennard-Jones and Coulomb interactions were calculated up to a distance of 1.2\,nm and long-range interactions for both were calculated using the Particle-Mesh-Ewald method \cite{Pedersen_JCP_1993_particle} and a Fourier-spacing of 0.144\,nm. The integration step was $\Delta t=0.2\,$fs.\\
Equilibration simulation runs lasted at least 12\,ns, extending up to 200\,ns at lower temperatures to ensure a minimum mean-square displacement of 50\,nm$^2$ across all temperatures. These were succeeded by 3.5\,ns sampling simulations at constant mean densities for each temperature or until achieving a 20$\,$nm$^2$ displacement, from which ten time-equidistant configurations with instantaneous velocities were selected as initial setups for 0.5\,ns production runs, recorded every 2\,fs. This approach allowed sampling from distinct micro-states and enabled the capture of higher frequency dynamics, averaging the outcomes across all production runs per temperature for the following analysis.\\
For the calculation of the IR spectra from these simulations, the classical representation with the Fourier transform of the autocorrelation of the total dipole moment was employed \cite{mcquarrie2000time}
\begin{equation}\label{eq_dipol1}
    I(\omega) \sim \int_{-\infty}^\infty dt e^{-i\omega t} \left\langle \vec M(0)\cdot\vec M(t) \right \rangle \;.
\end{equation}
where $\vec M=\sum_i \vec \mu_i$ is the total dipole moment of the system and the angular brackets $\langle\cdot \rangle$ denote an ensemble average and averages over several time origins. The self-correlation spectra are obtained by taking only the summation $\sum_{i,j=i} \vec \mu(0) \cdot \vec \mu(t)$, i.e.
\begin{equation}\label{eq_dipol2}
    I_\mathrm{self}(\omega) \sim \int_{-\infty}^\infty dt e^{-i\omega t} \left\langle \vec \mu_i(0) \cdot \vec \mu_i(t) \right \rangle\;.
\end{equation}
The spectra were divided by temperature to compensate its effect on the intensity of the absorption.

\section{Results}

\subsection*{Experimental FTIR spectra of H$_2$O}
\begin{figure}
\centering
\includegraphics[width=8.5cm]{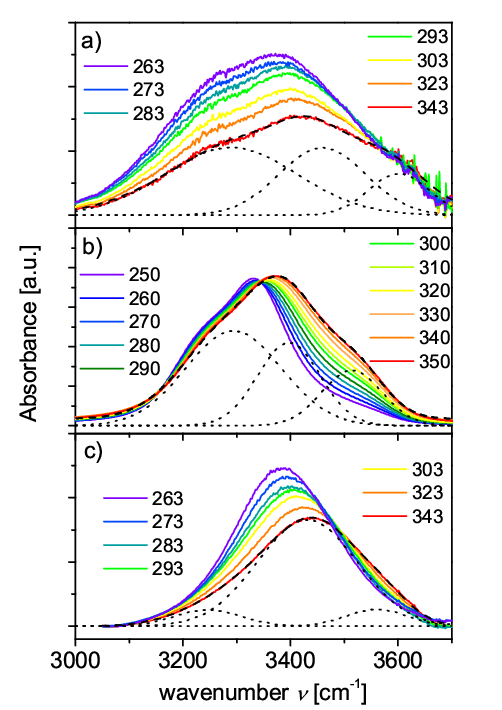}
\caption{Vibrational spectra in the OH-stretching range: a) experimental IR absorption spectra of pure H$_2$O, b) dipole self-correlation spectra of pure H$_2$O obtained from molecular dynamics simulation, and c) 2\% HDO in D$_2$O vibrations at different temperatures as indicated (in Kelvins). The dashed lines are exemplary fits to the spectra at the highest temperature in each panel composed of 3 Gaussians; these individual Gaussian contributions are shown separately as dotted lines.}
\label{IR-spectra}
\end{figure}

The IR absorption spectra of H$_2$O contain a broad structured peak in the range between $3000 - 3700\;cm^{-1}$ which is generally assigned to stretching vibrations of OH groups with the portion below $\sim 3600\;cm^{-1}$ reflecting those involved in inter-molecular H-bonds (Fig. \ref{IR-spectra} a). As temperature decreases the overall absorbance of this peak increases and its maximum position at $\sim 3400\,$cm$^{-1}$ as well as its shoulders at $\sim 3300\,$cm$^{-1}$ and slightly below $3600\,$cm$^{-1}$ shift to lower wavenumbers. This indicates a weakening of the covalent OH bond which is typical for inter-molecular H-bond forming materials since the increase in density brings the H-bond acceptor closer to the OH, thus stretching the interatomic potential of the covalent OH-bond.\\
According to the literature, the assignment of the various contributions of the structured OH stretching band is ambiguous and has been controversially debated \cite{Riemenschneider2009,Perakis2016}. Thereby, two opposing ideas appear to dominate: in one view, individual peaks of the complex band are considered to reflect states of a specific number of active donor and acceptor sites in the probed molecule \cite{Riemenschneider2009} or to symmetric and antisymmetric stretching modes \cite{DeMarco2016}, thus expecting a superposition of several Gaussian peaks reflecting the number of states \cite{Max2002,Sun2009,Kataoka2011,Max2015,Max2017}; the other interpretation distinguishes contributions originating either from isolated or from collective vibrations, the latter ones are sometimes regarded cross-correlations, combination bands or coupled vibrations in which case neither the precise number nor the exact shape is obvious \cite{Choi2013,Nagata2015,Medders2015,Perakis2016,Carlson2020}. The latter view has also been established in the terahertz range of the H$_2$O absorption \cite{Sharma2023}. Essential for the present investigation is to unravel the contributions and identify which ones originate from isolated vibrations and thus can be reasonably related to physical properties of individual bonds, e.g. its length.\\

\subsection*{Dipole correlation spectra of H$_2$O from MD simulations}

Since the absorption of IR radiation is based on the presence of a transition dipole moment of matching transition energy, the dipole correlation spectra of water were deduced from MD simulations (Fig. \ref{IR-spectra}b). Thereby, the self-correlation spectra of the molecules exhibit a remarkable resemblance with the experimental results, also consisting of three major contributions with a prominent peak in the center at about $\sim3400\;cm^{-1}$ which shifts significantly to lower wavenumbers as temperature decreases, a pronounced low-frequency-shoulder at $\sim3300\;cm^{-1}$, and a weaker shoulder at about $\sim3550\;cm^{-1}$ which increases in intensity as temperature rises. Although the experimental FTIR spectra are considerably broader, a feature which is often observed in simulations \cite{Zhong2020}, the peak positions of the three major contributions are quite similar.\\
Despite the fact that only the self-correlations were calculated, these simulated spectra are an intricate mixture of isolated and combined vibrations because a highly coupled system is modeled. The similarity to the experimental FTIR spectra suggests that also the latter contain a considerable contribution of collective vibrations (intra- as well as inter-molecular) which is not easily disentangled.\\

\subsection*{Experimental FTIR spectra of dilute HDO in D$_2$O}

An experimental way to largely isolate the OH bond vibration, i.e. to minimize combination bands of collective vibrations, is to dissolve HDO in D$_2$O \cite{Marti1994,DeMarco2013}. We examine a volume concentration of $2\%$ HDO in D$_2$O which means that statistically there is one OH bond per 50 water molecules. Assuming an even distribution this separates OH bonds by $\sim 1.3$\,nm which should strongly reduce energy transfer between these oscillators.  The respective IR spectra exhibit a much narrower peak at $\sim 3400\,$cm$^{-1}$ than in the case of H$_2$O (though it is still broad compared to other absorption bands) with very little structure; only a slight asymmetry indicates the presence of additional contributions at lower and higher frequencies (figure \ref{IR-spectra} c). This confirms that the spectrum of H$_2$O indeed contains several contributions of combined vibrations and a single isolated vibrational peak.\\
Further evidence for this interpretation is provided by the rather unstructured peak shape one can observe in materials with lower concentrations of OH bonds, e.g. polyalcohols \cite{gabriel2021}: despite the fact that the formation of up to three inter-molecular H-bonds per oxygen (O) is possible (i.e. different states of coordination are possible), no additional shoulders are found. The spatial separation of the OH groups due to the attachment to a carbon backbone in these materials strongly reduces the correlation among the vibrations of the OH bonds.\\

\subsection*{Quantitative analysis of vibrational spectra}

\begin{figure}
\centering
\includegraphics[width=8.5cm]{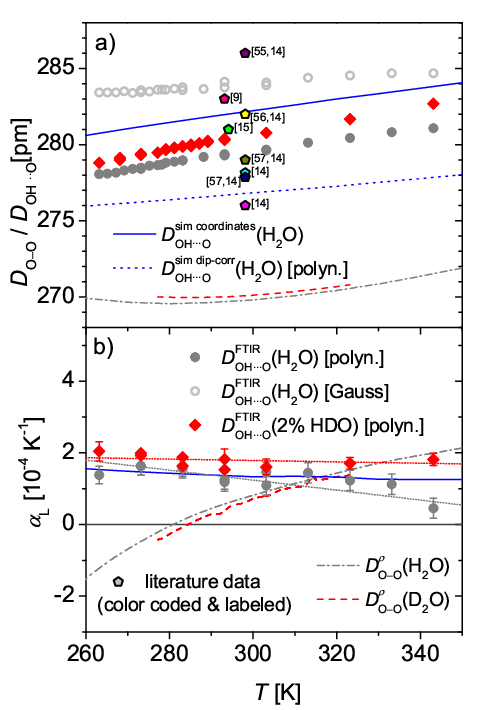}
\caption{a) Inter-molecular H-bond length as calculated via equ.\,\ref{eq:ohofunction} from the frequency of the main Gaussian peak (open grey circles) as well as the global maximum deduced with a polynomial fit (solid grey circles) of the FTIR spectra of H$_2$O and 2\% HDO in D$_2$O (solid red diamonds), respectively, as well as the average H-bond length in the MD simulations of H$_2$O (solid blue line). The dotted line corresponds to the converted maxima of dipole correlation peaks from the simulation deduced via a polynomial fit and equ.\,\ref{eq:ohofunction}. Additionally, several literature values of the first peak in the radial O-O pair correlation function deduced either from X-ray scattering \cite{Bergmann2007,Gorbaty1985} or from a combined analysis \cite{Soper2007} of several older X-ray \cite{Narten1971,Hura2003,Hart2005} and neutron scattering \cite{Soper2005,Soper2000} data sets are shown. b) Deduced linear expansion coefficient $\alpha_L$ of select data shown in a) using the same symbols and color code; dotted red and grey lines are linear fits to the data of HDO in D$_2$O and H$_2$O, respectively. In order to reduce the scattering in the numerical differentiation, only data points in $10\,\text{K}$ steps were incuded in b). The experimental uncertainty in panel a) is smaller than the symbol size; in panel b) the uncertainty is displayed for some representative data points for clarity.
}
\label{IR-lengths}
\end{figure}

Despite past approaches to fit the OH-stretching region of H$_2$O with five or more Gaussians accounting for the potential number of donors and acceptors or other assumed molecular states \cite{Max2002,Sun2009,Kataoka2011,Max2015,Max2017}, a sum of three Gaussian functions appears to be sufficiently accurate to fit the measured and simulated spectra of H$_2$O and HDO in D$_2$O to extract peak position, width, and area of the individual contributions (figure \ref{IR-spectra}). However, the close superposition of the contributions leads to a strong dependence on these fit parameters and thus large uncertainties. Consequently, we also fitted a polynomial curve to the spectra of HDO in D$_2$O to deduce the position of the maximum with much better precision (which due to the suppressed contributions of collective vibrations, in this case, corresponds to the peak of the vibration of the individual bond).\\
To improve also the extraction of the width and area of the peaks in the spectra of HDO in D$_2$O, the fit with three Gaussian functions was repeated with the maximum position of the central peak fixed to the value obtained from the polynomial fits. Finally, also the experimental H$_2$O spectra were fitted again with three Gaussian functions with the width of the central peak fixed to the value obtained from the improved fit of the spectra of HDO in D$_2$O.\\
To deduce inter-molecular bond lengths, the peak position of the isolated vibration, i.e. the maximum at $\sim 3400\,$cm$^{-1}$, of both systems is converted to the corresponding OH$\cdots$O length $D_{\text{OH}\cdots\text{O}}$ via an empirical relation that describes the correlation observed in a large data set of several H-bonding crystalline materials \cite{Libowitzky1999,steiner2002}
\begin{equation}
 D^{\text{FTIR}}_{\text{OH}\cdots\text{O}}(\bar \nu) = 13.21 \log \left( \frac{304\cdot10^{9}}{3592-\bar \nu\cdot \mathrm{cm}} \right) \mathrm{pm}
 \label{eq:ohofunction}
 \end{equation}
where $\bar \nu$ is the peak position in cm$^{-1}$ and the result is obtained in picometers ($\text{pm}$). A similar approach has been taken in the past based on Raman spectra \cite{Tulk1998}. The extracted length is in the range of $280\,$pm which is typical for the OH$\cdots$O distance in inter-molecular H-bonds. Further, as expected for inter-molecular bonds the length expands with increasing temperature for both pure H$_2$O and dilute HDO in D$_2$O (fig. \ref{IR-lengths} a).\\ 
Especially the latter match well with the average OH$\cdots$O distance along the inter-molecular H-bonds deduced from the coordinates of the molecules in the simulation. In contrast, an attempt to convert the frequency positions of the central peak of the simulated dipole correlation spectra through equ.\,\ref{eq:ohofunction} yields a length that is about $5\,$pm smaller and, more importantly, much less steep. The latter demonstrates the limitations of the simulation to capture details like the correct temperature dependence of the relation between inter-molecular distance and respective vibrational absorption.

\subsection*{Crosscheck with mass density}

In the past, several studies established a conversion from mass density $\rho$ to the inter-molecular distance (even more specifically the separate lengths of covalent and inter-molecular H-bond were presented) \cite{Huang2013,Sun2023,Zhou2022}. This is certainly motivated by the simple internal structure of the water molecule which appears promising (if not mandatory) for linking the inter-molecular length scale with a macroscopic quantity as attempted in the present study. Thereby, the inter-molecular H-bond length was considered to correspond to the average distance of nearest-neighbor pairs of oxygen atoms $D_{\text{O-O}}$ which for H$_2$O was estimated as \cite{Huang2013,Sun2023,Zhou2022}
\begin{equation}
    D^{\rho}_{\text{O-O}}(\text{H}_2\text{O})=2.695\times10^{-7} \text{g}^{1/3} \rho^{-1/3}
    \label{equ:volumes1}
\end{equation}
where the prefactor carries the cube root of grams $\text{g}^{1/3}$ as unit. This equation should also be applicable to D$_2$O since it is governed by similar inter-molecular interactions; to account for the difference in molecular mass, a correction factor of $\left(\frac{M_{\text{D}_2\text{O}}}{M_{\text{H}_2\text{O}}}\right)^{1/3}$ must be introduced
\begin{equation}
    D^{\rho}_{\text{O-O}}(\text{D}_2\text{O})= 2.791\times10^{-7} \text{g}^{1/3} \rho^{-1/3}
    \label{equ:volumes2}
\end{equation}
However, the thus obtained lengths are in the range of $\sim270\,\text{pm}$ for both H$_2$O and D$_2$O, i.e. about $\sim10\,\text{pm}$ shorter than we estimate for the length of the inter-molecular H-bonds from the FTIR data (Fig.\,\ref{IR-lengths} a). Furthermore, the temperature dependence is quite different being non-monotonous with minima at $277\,K$ and $284\,K$, respectively, which indicate the density maxima of both materials \cite{Herrig2018}, in contrast to the monotonous increase of $D_{\text{OH}\cdots\text{O}}$ deduced from FTIR. Consequently, the direct conversion of mass density into inter-molecular H-bond length (or intra-molecular ones for that matter) \cite{Huang2013,Sun2023,Zhou2022} is incorrect.\\

\section{Discussion}

\subsection*{Sphere model}

As is well-known, the unusal thermal evolution of the mass density of water is governed not simply by inter-molecular bond expansion but also by a complex structural reorganization \cite{MallamaceF2020}. To take this into account, we set up a slightly more complex, yet still simplistic geometric model that approximates the water molecules as spheres, an approach which is supported by X-ray scattering results \cite{Narten2003}. Thereby, the oxygen atom is in the center of the sphere with a radius equal to half the length of an inter-molecular H-bond (i.e. $r_s=0.5\cdot D_{\text{OH}\cdots\text{O}}$) and the hydrogen atoms are located roughly at the surface of this sphere corresponding to two of the edges of an inscribed tetrahedron (actually, the surface of the sphere would cut the inter-molecular H-bond in half which means that the covalently bound hydrogen atoms are considerably closer to the center).\\
In this picture, liquid water resembles a loose packing of these spheres with a coordination number of $Z\leq4$; thereby, lower coordination numbers account for the time average of broken inter-molecular H-bonds during the rearrangement of the network. (We would like to note that a random close packing of spheres corresponds to $\sim$6 nearest neighbors, i.e. neighbors with direct contact; hence, an equivalent of a random loose packing has to be employed in our model. In simulations, random loose packing is achieved by considering spheres with surface friction \cite{Liu2015} - not unlike the concept of molecular friction in liquids and glasses.) In a random loose packing, the number of nearest neighbors $Z$, i.e. the coordination number, is related to the volumetric space filling $f_s$ of the spheres \cite{Liu2015}. With the latter, we can relate the volume occupied by the spheres $V_s$ to the total volume $V_t$ as $V_s=f_s V_t$. Then, $V_s$ can be replaced by the volume of a single sphere and $V_t$ by the specific volume given by mass density $\rho$ and molecular mass $M_W$ which yields
\begin{equation}
    \frac{4}{3}\pi r_s^3=f_s \frac{M_W}{\rho N_A}
    \label{equ:volumes3}
\end{equation}
Consequently, the space-filling of the equivalent spheres can be estimated from the average distance between oxygen atoms of adjacent water molecules $D_{\text{OH}\cdots\text{O}}=2r_s$ as
\begin{equation}
    f_s =  \frac{\pi\rho N_A}{6 M_W} {D_{\text{OH}\cdots\text{O}}}^3
    \label{equ:density-length}
\end{equation}
With an empirical relation $Z(f_s)=2+11f_s^2$ taken from the literature \cite{German2014} the respective average sphere-sphere coordination numbers are obtained according to
\begin{equation}
    Z =  2+11\left(\frac{\pi\rho N_A}{6 M_W}\right)^2 {D_{\text{OH}\cdots\text{O}}}^6
    \label{equ:coord-length}
\end{equation}
The resulting values are in the range of 3.5 to 3.6 for both H$_2$O and dilute HDO in D$_2$O. Although respective values in the literature range from $\sim2-4$ \cite{Kontogeorgis2022,Head-Gordon2006}, the standard view seems to consider $3.5-3.6$ H-bonds per molecule in liquid water \cite{Liu2021a}. E.g. some X-ray diffraction data shows $4.4$ neighbors in the first shell, though a considerable proportion of these are more than $3.3\,\text{\AA}$ apart, which means these are outside the typically considered cut-off length of inter-molecular H-bonds \cite{Gorbaty1985}. Using the cut-off criterion yields $\sim3.5$ neighbors in the $2.8-3.3\,\text{\AA}$ range \cite{Gorbaty1985,Liu2021a}. Hence, the values obtained via our simplistic model are reasonable and thus validate the employed approach to quantitatively extract inter-molecular distances from FTIR bands.\\

It is important to note that the used equivalent sphere model intrinsically implies straight inter-molecular H-bonds while in reality these can be established over quite a range of angles \cite{steiner2002,Sun2023}.\\
Assuming that the frequency position of the OH stretching vibration is only sensitive to the proximity of an H-bond acceptor but not to the distance, the extracted $D_{\text{OH}\cdots\text{O}}$ may be overestimating the true distance depending on the angle. Inter-molecular H-bonds are established under angular deviations of up to $20^\circ$ from the straight arrangement of oxygen-hydrogen-oxygen atoms \cite{Steiner1994}. In this extreme case, the true value is reduced by a factor of $\sim0.986$, i.e. $1.4\%$ shorter (pure geometric estimation using the length ratio $\varphi$ between the covalent and the inter-molecular part of the OH$\cdots$O-bond and the angular deviation $\theta$ yield the equation for this correction factor $R=\frac{\varphi\,\text{cos}(\alpha)+\text{cos}(\theta-\alpha)}{\varphi+1}$ with $\alpha=\text{arctan}\left(\frac{\text{sin}\theta}{\varphi+1-2\,\text{sin}^2\left(\theta/2\right)}\right)$). Inserting this in equation \ref{equ:coord-length} yields a systematic overestimation of $Z$ of less than $4\%$ for H-bonds with an angle of $160^\circ$.\\
Further, the equivalent sphere model also does not take into account the tetrahedral arrangement of H-bonds around one oxygen atom. Instead, in the model, any sufficiently close adjacent sphere contributes to $Z$ even if in reality no inter-molecular H-bond can be established (e.g. if the central oxygen atom has already formed four H-bonds or if no tetrahedral arrangement with the other H-bonds is achieved). Consequently, also this will lead to an artifactual increase in the estimated coordination number. Considering these intrinsic inaccuracies of this rather simple model the close match with the literature appears remarkable.\\

\subsection*{Thermal expansion}

\begin{figure}[hbt]
\centering
\includegraphics[width=8.5cm]{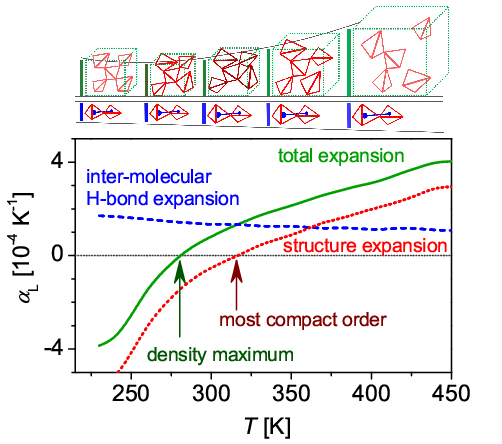}
\caption{Schematic plot of the linear thermal expansion coefficient $\alpha_L$ of water and its separated components, i.e. the thermal expansion of the inter-molecular H-bonds and the structural reorganization. The latter crosses from positive to negative values at a temperature in the range $316-331\,\text{K}$ which indicates the most compact packing of the molecules. However, the nearly constant and positive expansion of the inter-molecular H-bonds shifts this cross-over in the global expansivity to lower temperatures where it is observed as maximum in the mass density. The sketch above displays schematically how the parallel evolution of steadily increasing inter-molecular distance (visualized as size of the tetrahedrons and the blue bar) and the non-monotonous structural expansion (red) in a tetrahedral arrangement combine to the total specific volume (green boxes and bars); at low temperatures perfect tetrahedral order is achieved which contains a lot "free space", at moderate temperatures the arrangement is irregular but most compact, and at high temperatures an irregular arrangement with increased "free space" is generated due to reduction of nearest-neighbor contacts. Note that the smallest volume, i.e. highest density is not achieved with the most compact arrangement due to the steady expansion of the molecules.}
\label{scheme}
\end{figure}

Although the sphere model reasonably links the inter-molecular H-bond length with the macroscopic mass density, the temperature dependencies of the respective lengths exhibit a qualitative difference (Fig.\,\ref{IR-lengths} a). This emphasizes a fundamental difference between these two quantities. To examine this in more detail we calculate the linear expansion coefficient $\alpha_L$ via the numeric derivative of the H-bond length with respect to temperature and, in the case of mass density, via the volumetric expansion coefficient $\alpha_V$ and the relation $\alpha_V=3\alpha_L$ (Fig.\,\ref{IR-lengths} b). 
The inter-molecular H-bond length shows a weakly decreasing expansion coefficient of about $1.7-1.9\times10^{-4}\,\text{K}^{-1}$ in the measurements on dilute HDO, and $1.3-1.5\times10^{-4}\,\text{K}^{-1}$ in the simulations of H$_2$O. In slight contrast, the experimental data of H$_2$O exhibit a considerable reduction from $1.8\times10^{-4}\,\text{K}^{-1}$ to $0.5\times10^{-4}\,\text{K}^{-1}$ which may suggest that the contribution of collective bands in the respective FTIR spectra is too intense to reasonably extract minute characteristics of the isolated vibration. However, While in all these cases the inter-molecular H-bond length shows a positive expansion coefficient in the range $0.5-1.9\times10^{-4}\,\text{K}^{-1}$, the mass density shows values varying from $-1.5\times10^{-4}\,\text{K}^{-1}$ at $260\,K$ to $2\times10^{-4}\,\text{K}^{-1}$ at $350\,K$.\\
It is well-known that the thermal expansion of water is dominated by orientational effects that enable certain packing structures rather than solely the thermally controlled position in the potential of inter-molecular potentials. However, a close inspection of the inter-molecular H-bond length reveals that the latter effect also contributes significantly to the macroscopic expansion (Fig.\,\ref{IR-lengths} b). In the context of supra-molecular arrangement the expansion of the inter-molecular bond length can be viewed as the effective thermal expansion of the molecule (used here in the sense of the constitutional particles of the liquid), hence, subtracting this portion from the macroscopic expansion should yield the contribution of the structural reorganization. Consequently, one would expect the intersection point of the macroscopic expansion coefficient and the one of the inter-molecular H-bonds ($\sim331\,\text{K}$ using the $D_{\text{OH}\cdots\text{O}}$ deduced from dilute HDO or $\sim316\,\text{K}$ from the simulated $D_{\text{OH}\cdots\text{O}}$) to indicate the temperature of most compact structural order if the inter-molecular bond length was temperature independent (Fig.\,\ref{scheme}).\\
In fact, at about this temperature the several pressure-related anomalies can be observed in water: the compressibility exhibits a minimum ($\sim320\,\text{K}$) \cite{Kell1975}, the thermal expansivity is invariant of pressure ($\sim316\,\text{K}$)\cite{Mallamace2013,MallamaceD2020,MallamaceF2020} as is the difference between the heat capacities at constant pressure and constant volume $C_p-C_v$ ($\sim313\pm4\,\text{K}$) \cite{MallamaceF2020}, the isothermal piezo-optic coefficient (the derivative of the refractive index with respect to pressure) has a minimum ($\sim318\,\text{K}$) \cite{Shraiber1975}, the heat capacity at constant pressure $C_p$ shows the smallest variation with pressure ($\sim325\pm10\,\text{K}$) \cite{Mallamace2013}, and the pressure dependence of the zero shear viscosity changes from monotonous increase to a curve with a minimum ($\sim307\,\text{K}$) \cite{Bett1965}. It has been proposed earlier that the minimized susceptibility to pressure change revealed in these anomalies indicates the establishment of the most compact supra-molecular arrangement rather at the temperature of $T^*\approx315\,\text{K}$ than at the actual density maximum \cite{Mallamace2012,Mallamace2013,MallamaceD2020,MallamaceF2020}.\\

\subsection*{Isotope effect}

A remarkable aspect thereby is the gradual increase of the temperature at which the density maximum occurs as the atomic mass of the involved hydrogen isotope increases; for D$_2$O it is at $\sim284\,\text{K}$ \cite{Herrig2018}, and for T$_2$O even $\sim287\,\text{K}$ \cite{Goldblatt1964}. Considering that a higher mass of the hydrogen isotope results in a stronger covalent bondwith the oxygen atom a reduced thermal expansivity of that can be expected. With the simplistic assumption of the same contribution of the structural rearrangement to thermal expansivity, one can indeed explain an increased temperature of the cross-over point from negative to positive thermal expansivity.\\
However, experimental studies show that in D$_2$O the inter-molecular H-bonds are stronger than in H$_2$O \cite{Nemethy1964}. This suggests also a higher stability of the supra-molecular structure which, on the one hand, would give rise to an additional increase of the cross-over temperature. On the other hand, stronger inter-molecular bonds are less susceptible to a temperature which would lead to a shallower slope of the thermal expansivity. Then, the sum with the expansivity of the bond would shift stronger to lower temperatures, countering the other effects. Consequently, a qualitative assessment is insufficient to draw a final conclusion. Instead, quantitative investigations on D$_2$O similar to the one presented here for H$_2$O (including a correlation of the DO stretching vibration frequencies with the respective inter-molecular bond length) are required to resolve this question.\\

\subsection*{Relation to X-ray and neutron scattering data}

The determination of the inter-molecular H-bond length in water has been the subject of several studies that employed X-ray or neutron scattering \cite{Gorbaty1985,Narten2003,Head-Gordon2006,Fu2009,Naberukhin2017,Krauss2019}, particularly by determination of the radial distribution function of oxygen atoms. The reported values for the average length of the nearest neighbor O-O distance is well within the range of our data, though the literature results exhibit quite some scattering (Fig.\,\ref{IR-lengths} a). While the data reported by Bergmann et al. \cite{Bergmann2007} and Soper et al. \cite{Soper2007} are a near perfect match with our results, others are larger or smaller by up to $5\,\text{pm}$. However, it should be noted that due to negligence of the angular asymmetry of the H-bond, our results may systematically overestimate the length by up to $1.4\%$ or $\sim4\,\text{pm}$, i.e. putting them potentially in agreement with some of the lower literature values. Although all of these differences correspond to less than $\pm 2\%$ around the $281\,\text{pm}$-mark, this deviation is bigger than the thermal expansion we find for the inter-molecular H-bond in the temperature interval $260-350\,\text{K}$ (expansion by $\sim3\,\text{pm}$ or $\sim1\%$).\\
We would like to emphasize that water is a special case where, due to the simple structure of the individual molecule, the inter-molecular H-bond length can be easily evaluated from scattering methods probing the oxygen pair correlation function. However, for more complex molecules that engage in H-bonding, this is far more demanding - especially if intra-molecular conformational changes are possible. Particularly for these latter cases, the here presented analysis of vibrational spectroscopy data is a valuable addition to determine specifically the inter-molecular H-bond length. As has been demonstrated, the separate estimation of this quantity combined with information from the macroscopically deduced specific volume enables a better understanding of molecular packing and structural changes.\\

\subsection*{Implications for the IR spectra of water}

Some of the relations and findings discussed above have notable consequences for the expected shape of the IR absorption band of the OH stretching vibration. Based on the correlation between the inter-molecular H-bond length and the peak position of the OH stretching band, one can conclude that the spectral shape of this band reflects the distribution of bond lengths in the system. Thereby, two opposing effects complicate the conversion. First, the nearest neighbor oxygen pair correlation function exhibits an asymmetry along the radial coordinate in the probability distribution of the first coordination shell \cite{Fu2009}. Particularly, the distribution is stretched towards longer distances demonstrated by the fact that the mean distance of the nearest neighbor is larger than the most probable distance. This would result in a skewed IR absorption band with an extended high-frequency flank. However, the conversion according to equ.\,\ref{eq:ohofunction} introduces another distortion due to its nonlinear shape. In this case, larger H-bond lengths are mapped onto a narrower frequency interval than short ones which would result in a stretched low-frequency flank of reduced intensities while the high-frequency flank would exhibit a compression in frequency but an increase in intensity. Although these two contributions have opposite effects it is unlikely that they cancel out. Consequently, a general expectation for the broad absorption band of the OH stretching vibration is, even without considering the sidebands from collective vibrations, an asymmetric shape. As a result, a fit to one or more Gaussian functions is unjustified not only from a physical point of view but it may also deliver incorrect values of the peak position, as was demonstrated in the present investigation for the H$_2$O spectra.\\

\section{Conclusion}

In this study we use a previously established correlation between the IR absorption frequency of the OH-stretching vibration and the corresponding length of the inter-molecular H-bond \cite{Libowitzky1999,steiner2002} in order to extract the average inter-molecular distance in liquid water in a wide temperature range. For that, the highly structured absorption band of the OH-stretching vibration was unraveled, and several contributions were assigned: the central and most pronounced peak was found to reflect the isolated OH-bond vibration ($\sim 3400\,$cm$^{-1}$) and two additional shoulders resemble cooperative vibrations of this highly coupled system of oscillators ($\sim 3250\,$cm$^{-1}$ and $\sim 3550\,$cm$^{-1}$). Although alternative views exist which assign each absorption peak to a specific state of a particular number of donors and acceptors of inter-molecular H-bonds \cite{Max2002,Sun2009,Kataoka2011,Max2015,Max2017}, several studies interpret considerable contributions to correspond to collective vibrations and combination bands \cite{Choi2013,Nagata2015,Medders2015,Perakis2016,Carlson2020}. Our experimental data in a system of dilute OH-bonds (i.e. 2$\%$ HDO in D$_2$O) confirms the latter view by showing a much narrower peak with significantly reduced contributions of these collective vibrations. This should lead to rethinking some of the previous interpretations of IR spectra of water.\\
The obtained inter-molecular H-bond length of about $\sim283\,\text{pm}$ corresponds well with values reported by past studies using X-ray and neutron scattering \cite{Bergmann2007,Soper2007,Gorbaty1985}, although the scattering of these and other literature results exceeds the thermal expansion we observe in the investigated temperature interval ($260-350\,\text{K}$). A commonly used approach to calculate the average nearest neighbor pair distance of oxygen atoms from the mass density yields by about $10\,\text{pm}$ lower values and, more importantly, a significantly different temperature dependence, e.g. a clear minimum corresponding to the density maximum at $277\,\text{K}$ in H$_2$O ($284\,\text{K}$ in D$_2$O). In contrast, the inter-molecular H-bond length obtained from the IR data is monotonously increasing in the whole investigated temperature range.\\
Despite this discrepancy, the extracted inter-molecular H-bond length can be related to the macroscopic mass density through a simple equivalent sphere model with a variable coordination number to adjacent molecules. The resulting values of $\sim3.5-3.6$ nearest neighbors in direct contact are well within the expected range \cite{Gorbaty1985,Liu2021a}. This match verifies that the determination of inter-molecular bond lengths from IR absorption frequencies (via a calibrated relation\cite{steiner2002}) and yields accurate results in the case of inter-molecular H-bonds. However, the need for an additional degree of freedom (i.e. the variable coordination number) in the description demonstrates that the thermal expansion of water consists of two contributions: i) the structural arrangement of molecules with respect to their neighbors, and ii) the thermal expansion of the inter-molecular H-bond.\\
With experimental access to the latter and determination of the overall thermal expansion through mass density one can extract the purely structural contribution to thermal expansion. As a result, the most compact structure, i.e. considering only molecular arrangements but not the thermal expansion of inter-molecular bonds, is found at considerably higher temperatures than the density maximum, namely at $\sim331\,\text{K}$ from the experimental data and $\sim316\,\text{K}$ in the simulations. This assignment is corroborated by the coincidence with the temperature of several anomalies related to the susceptibility of the material to pressure; past studies have suggested a single characteristic temperature $T^*\approx315\,\text{K}$ for all these anomalies \cite{Mallamace2012,Mallamace2013,MallamaceD2020,MallamaceF2020}. Here, we would like to note that for the determination of $T^*$ through these anomalies exhaustive measurements in the temperature and pressure parameter space have to be carried out. In contrast, with the presented method it is possible to identify $T^*$ without varying the pressure.\\

For water the inter- and intra-molecular bond lengths (and also bond angles for that matter) and their thermal evolution are extremely well-studied via different spectroscopic and other methods \cite{Ford1968,Marechal1991,Corcelli2005,DeMarco2016,Zhou2022,Sun2023} as well as a multitude of simulations \cite{Marti1994,Corcelli2005,Choi2013,Nagata2015,Carlson2020,Liu2021a}. However, most other liquids lack such a profound set of data. Thus, we propose a combination of spectroscopic (e.g.FTIR) examination of inter-molecular bond lengths and density/specific volume measurements to disentangle different contributions to thermal expansivity and gain more insight into the correlation of structure and dynamics in disordered systems like liquids and glasses. Consequently, this method can pave the way to study inter-molecular distances and untangle structural effects - and due to the chemical specificity of IR spectroscopy even site-specific - in amorphous systems which are not easily addressed by conventional scattering methods (due to their need of long-range order to address specific bonds).\\

\section{Data availability}

The data sets generated and analyzed during this study are available from Jan P. Gabriel upon reasonable request.

\section{Acknowledgement}
Funding by the Deutsche Forschungsgemeinschaft (DFG, German Research Foundation) – Project-ID 189853844 – TRR 102 and by the VILLUM~Foundation \emph{Matter} grant (grant~no.~16515) is highly appreciated. We thank as well Michael Vogel for computational time.

\bibliographystyle{achemso} 
\bibliography{water-IR.bib}

\end{document}